\documentclass[journal]{IEEEtran} 

\IEEEoverridecommandlockouts                              

\usepackage[draft]{todonotes}   
\usepackage{soul}
\title{\LARGE \bf
SPARE: Spiking Neural Network Acceleration Using ROM-Embedded RAMs as In-Memory-Computation Primitives
}

\author{Amogh~Agrawal*, Aayush~Ankit* and Kaushik Roy,~\IEEEmembership{Fellow,~IEEE}

\thanks{All authors are with the School of Electrical and Computer Engineering, Purdue University, West Lafayette, US}
\thanks{Email: \{agrawa64, aankit, kaushik\}@purdue.edu}
\thanks{(* these authors contributed equally to this work)}
}

\usepackage{booktabs}
\usepackage{graphicx}
\usepackage{cite}

\begin{document}

\maketitle
\thispagestyle{empty}
\pagestyle{empty}


\begin{abstract}
From the little we know about the human brain, the inherent cognitive mechanism is very different from the \textit{de facto} state-of-the-art computing platforms. The human brain uses distributed, yet integrated memory and computation units, unlike the physically separate memory and computation cores in typical von Neumann architectures. Despite huge success of artificial intelligence, hardware systems running these algorithms consume orders of magnitude higher energy compared to the human brain, mainly due to heavy data movements between the memory unit and the computation cores. Spiking neural networks (SNNs) built using bio-plausible neuron and synaptic models have emerged as the power efficient choice for designing cognitive applications. These algorithms involve several lookup-table (LUT) based function evaluations such as high-order polynomials and transcendental functions for solving complex neuro-synaptic models, that typically require additional storage and thus, bigger memories. To that effect, we propose `SPARE' $-$ an in-memory, distributed processing architecture built on ROM-embedded RAM technology, for accelerating SNNs. ROM-embedded RAMs allow storage of LUTs (for neuro-synaptic models), embedded within a typical memory array, without additional area overhead. Our proposed architecture consists of a 2-D array of Processing Elements (PEs), wherein each PE has its own ROM-embedded RAM structure and executes part of the SNN computation. Since most of the computations (including multiple math-table evaluations) are done locally within each PE, unnecessary data transfers are restricted, thereby alleviating the problems arising due to physically separate remote memory unit and the computation core. SPARE thus leverages both, the hardware benefits of distributed, in-memory processing, and also the algorithmic benefits of SNNs. We evaluate SPARE for two different ROM-Embedded RAM structures $-$ CMOS based ROM-Embedded SRAMs (R-SRAMs) and STT-MRAM based ROM-Embedded MRAMs (R-MRAMs). Moreover, we analyze trade-offs in terms of energy, area and performance, for using the two technologies on a range of image classification benchmarks. Furthermore, we leverage the additional storage density to implement complex neuro-synaptic functionalities. This enhances the utility of the proposed architecture by provisioning implementation of any neuron/synaptic behavior as necessitated by the application. Our results show up-to $\sim1.75\times$, $\sim1.95\times$ and $\sim1.95\times$ improvement in energy, iso-storage area, and iso-area performance, respectively, by using neural network accelerators built on ROM-embedded RAM primitives.

\end{abstract}

\section{Introduction}

Deep Neural Networks (DNNs) are inspired from the hierarchical learning behavior in the human brain and have tremendously enhanced the learning capabilities in machines \cite{bengio2009learning,jones2014learning}. They have been credited to achieve high performance across a variety of recognition applications, even surpassing human abilities in certain tasks \cite{silver2016mastering}. In doing so, however, DNNs tend to consume orders of magnitude higher energy than the human brain. To bridge this energy gap, there have been proposals from the algorithm as well as hardware perspectives. Spiking neural networks (SNNs), or third generation neural networks, have evolved and have been shown to achieve comparable classification accuracies with respect to the non-spiking counterparts \cite{diehl2015fast}. SNNs rely on transfer of neuron spikes from one layer to the next, resembling the information transfer in the human brain. These spikes are encoded as binary data, thereby drastically simplifying the computations, and thus reducing the energy consumption.

On the other hand, hardware systems running DNN algorithms are inefficient, since DNN executions are memory- as well as compute-intensive. For instance, AlexNet which won the ImageNet 2011 challenge consists of 61 million parameters and involves 2-4 GOPS per classification \cite{krizhevsky2012imagenet,han2015learning}. Consequently, their execution on von-Neumann machines consumes more energy for data movement than computation \cite{han2015learning}. This can be attributed to the fact that DNN computation is inherently different from the conventional von-Neumann based computing model. Frequent data movement between a physically separate memory storage unit and a compute core forms the well known von-Neumann bottleneck. To overcome this bottleneck, there has been intense research for reducing data movements \cite{chen2017eyeriss}. Moreover, there have been proposals for \textit{in-memory computing} \cite{shanbhag,CRAM}, where the underlying principle is to perform the computations as close to the memory as possible, or better, within the memory array itself \cite{bor,resparc}.

Typical DNNs (with artificial neurons) involve multiple transcendental function evaluations (for instance, sigmoid, tanh, logarithms etc.). In addition, SNNs involve several bio-realistic neuron and synaptic differential equations, each having multiple transcendental function and high order polynomial evaluations. The most efficient way to implement such functions is by storing look-up tables (LUTs) and math tables in read-only memories (ROMs). However, large dedicated ROMs incur significant area and power overheads. To that effect, \cite{Lee2013} proposed embedding ROMs in standard CMOS SRAM caches (R-SRAM). R-SRAM allows placing a ROM within the conventional SRAM array (with corresponding architectural modifications), without degrading the area and performance benefits of the SRAM \cite{Lee2013}. Such compute primitives provide significantly higher storage densities (bits/area) which can be leveraged for DNN and SNN computations in storing useful data (LUTs) without affecting the RAM storage, thereby avoiding longer latencies and higher access energy associated with larger (or external) memory structures.


\begin{figure}[t]
\centering

\includegraphics[width=0.5\textwidth]{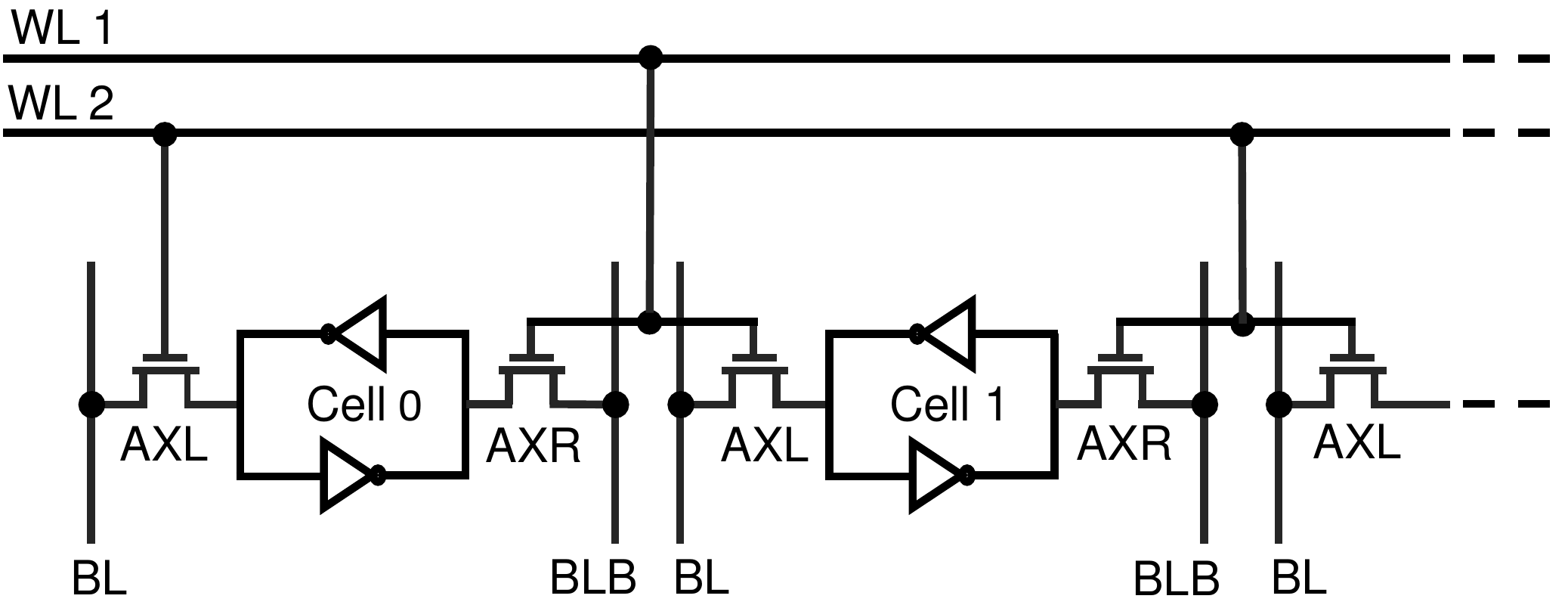}

\caption{R-SRAM Schematic: Standard 6T-SRAM embedded with ROM. The only difference is the addition of extra word-line (WL1 and WL2) to embed ROM functionality.}
\label{fig:emb}

\end{figure}

In this work, we take R-SRAMs and R-MRAMs a step further and propose ``SPARE'', a generalized architecture for SNN acceleration using ROM-embedded RAMs as in-memory-compute primitives. SPARE consists of a 2-D array of Processing Elements (PEs) that spatially map a deep SNN, where each PE performs part of the SNN computations. Each PE contains its own R-SRAM/R-MRAM which locally stores only the relevant synaptic data and the LUTs required for solving the neuron and synaptic differential equations. This localized processing leads to energy benefits, since only the neuron data (spikes) need to be transfered between PEs. Furthermore, since the PE operates only on an occurance of an input spiking event, unnecessary computations and memory accesses are avoided.
It is also worth noting that R-SRAM/R-MRAM primitive can store several different neuron and synapse models, thereby providing necessary flexibility. A PE thus, synergistically combines the hardware benefits from R-SRAMs/R-MRAMs and algorithmic benefits from SNNs. In summary, we make three key contributions.
\begin{enumerate}
\item \textbf{We design an energy-efficient PE} that leverages the ``in-memory processing'' abilities of ROM-Embedded RAM structures and ``event-driven computing'' in SNNs. We evaluate the pros and cons of using both, CMOS based R-SRAMs and STT-MRAM based R-MRAMs, as memory units in the PE.
\item \textbf{We design an efficient architecture (SPARE)} using a 2-D mesh of PEs, to provide a platform for cognitive application deployment. We show the implementation of spiking neural networks (fully connected and convolutional) on SPARE.
\item \textbf{We investigate the energy, performance and area benefits} for typical image classification benchmarks to underscore the system scalability and utility, both for training and inference phases.
\end{enumerate}

\begin{figure}[t]
\centering

\includegraphics[width=0.4\textwidth]{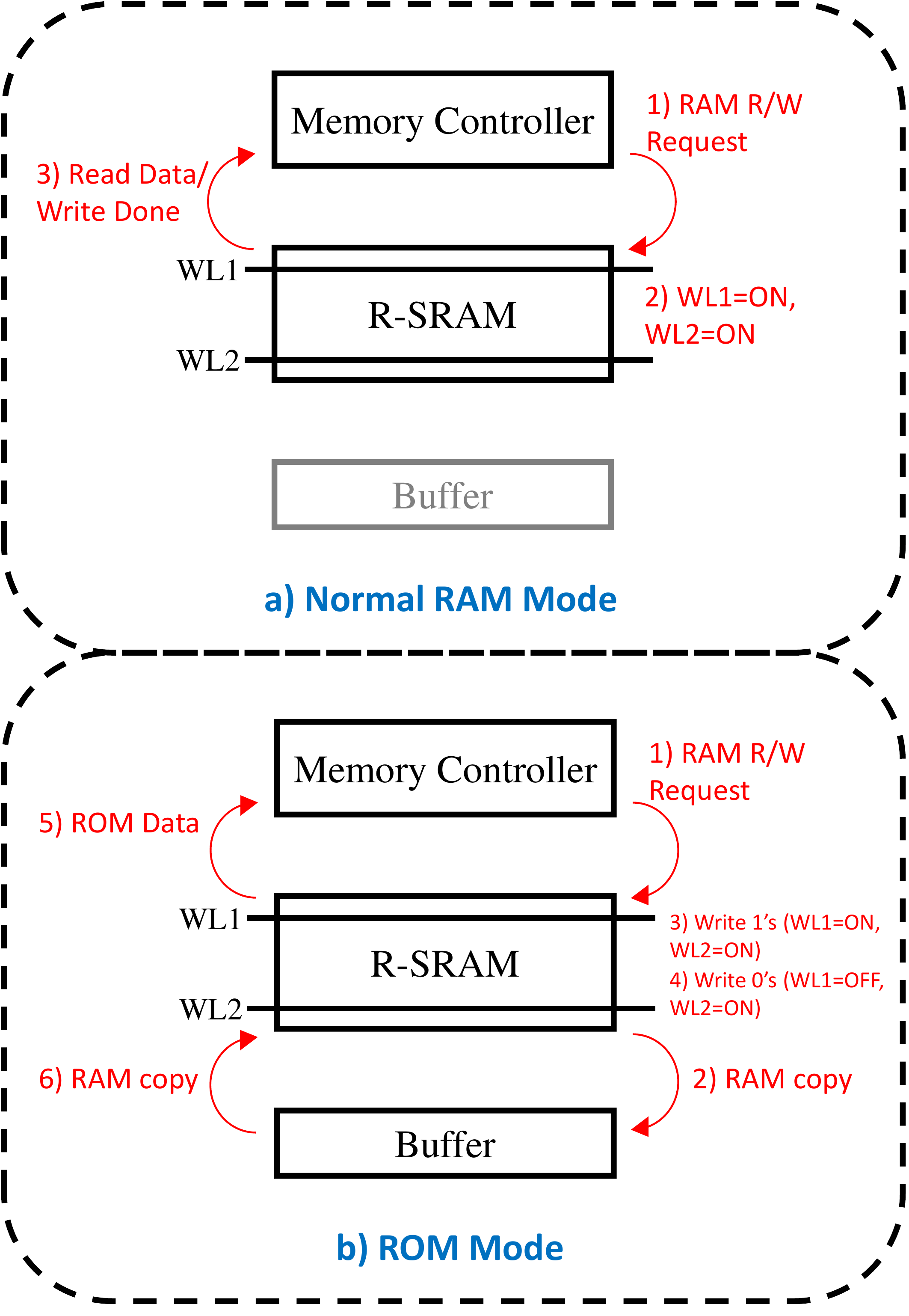}

\caption{Operation of R-SRAM in a) Normal RAM Mode and b) ROM Mode.}
\label{fig:roamoperation}

\end{figure}

\section{Background}

\subsection{ROM-Embedded RAMs}
\label{sec:roeam}
Previous studies on ROM-embedded RAMs were limited to logic testing and fast mathematical function evaluations \cite{Lee2013,fong2016embedding}. We explore the utility of R-SRAM and R-MRAM based memory structures towards designing efficient compute primitives for neuromorphic computing (SNN acceleration). Further, as discussed before, such memory units enable in-memory data processing that can be of immense utility in DNN execution, which are typically limited by the cost of data movements.

\paragraph{\bf{R-SRAM}}

R-SRAM is a memory structure that consists of a ROM in hardware embedded into a conventional CMOS SRAM array, with corresponding modifications at the architectural level to support ROM accesses \cite{Lee2013}. Fig. \ref{fig:emb} shows the structure of R-SRAM cell array \cite{Lee2013}. Unlike conventional 6T-SRAMs, R-SRAMs bit cells have an extra word-line (WL). The gate of the access transistors connect to WL1 or WL2, depending on the data to be embedded as ROM. Thus, if the bit-cell stores `0' (`1') as ROM data, the left access transistor
(AXL) is connected to WL2 (WL1). The right access transistor
(AXR) of the bit-cell follows the connectivity of the
AXL of the neighboring bit-cell to the right. For completeness, we describe the R-SRAM operation, both for the RAM mode and the ROM mode of operation.
\begin{enumerate}
\item \textit{RAM mode}: During the normal RAM mode, both word-lines, WL1 and WL2, are connected together. They are turned ON/OFF at the same time, so as to operate as conventional 6T-SRAM for memory read/write. Note that there is no performance penalty on RAM operations compared to the standard 6T-SRAM bit-cells.

\item \textit{ROM mode}: To retrieve the ROM data in the ROM mode of operation, a sequence of steps are performed, summarized in Fig. \ref{fig:roamoperation}. First, `1' is written to all bit-cells by turning both WL1 and WL2 ON. Thus, the whole row stores ``1111...''. Next, WL1 is turned OFF and `0' is written to all the cells, while WL2 remains ON. Now only the bit-cells connected to WL2 store `0', others store `1'. However, if two consecutive bit-cells have different ROM data, this step performs a 5T write operation on the SRAM cell, since only one access transistor is ON. This may lead to a ``write stability" problem in the bit-cells, which can be resolved using write-boost techniques \cite{Lee2013}. The ROM data can now be read using conventional RAM read operation. Note that the ROM data retrieval process destroys the initial RAM content. Hence, before ROM data retrieval, RAM data of the corresponding block is written into a buffer, as shown in Fig. \ref{fig:roamoperation}. After the ROM data has been retrieved, the RAM data of the block is restored.
\end{enumerate}

\begin{figure}[t]
\centering

\includegraphics[width=0.5\textwidth]{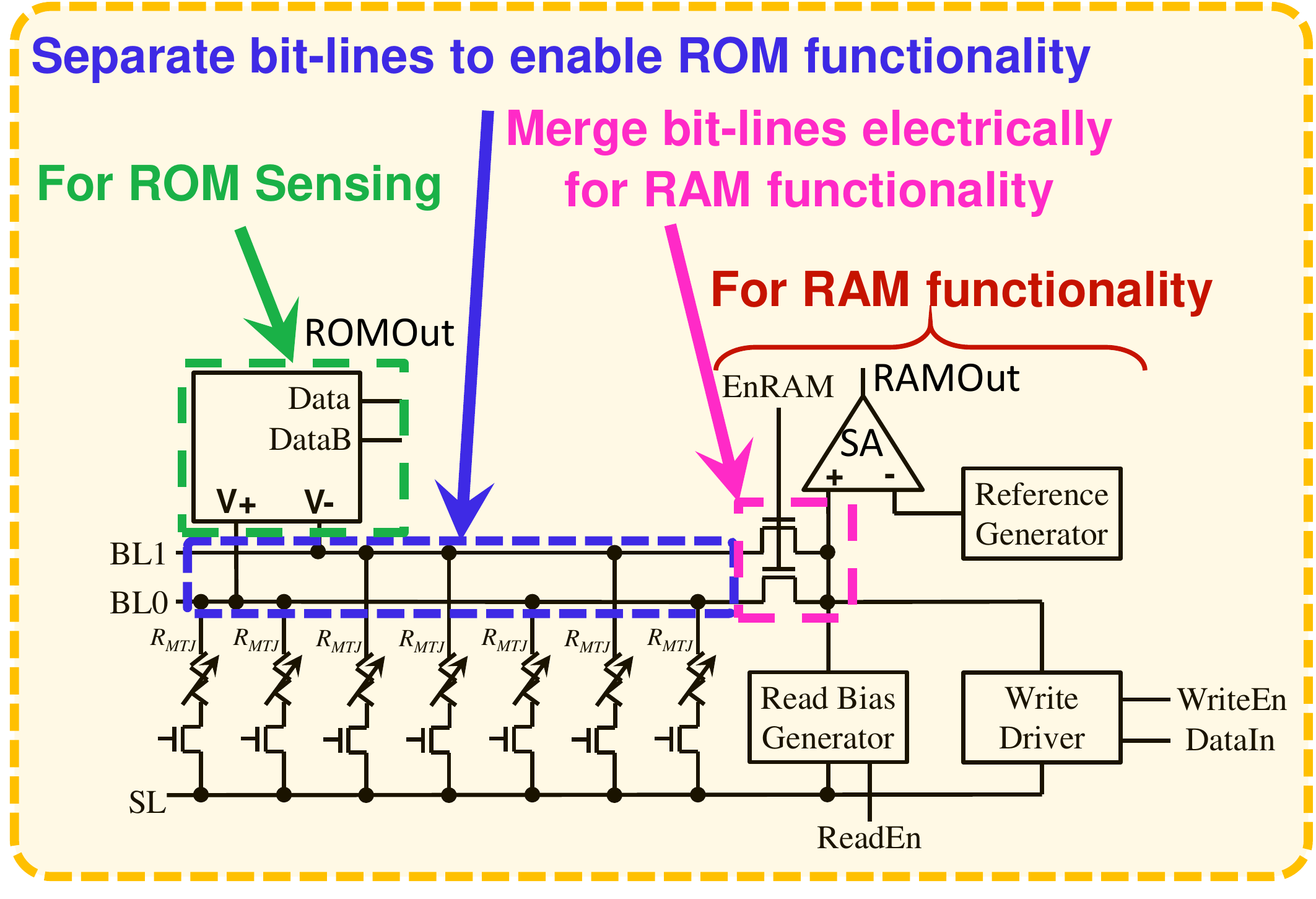}

\caption{R-MRAM Schematic: Standard STT-MRAM array with two bit-lines (BL1 and BL0) to embed ROM functionality. The peripheral circuitry for RAM and ROM mode of operation is highlighted.}
\label{fig:rmaram}

\end{figure}

It has been shown that R-SRAM incurs insignificant area ($\sim2\%$) and power ($\sim1\%$) overheads \cite{Lee2013} to incorporate an additional word-line requirement. Moreover, we will show later in our simulations that despite the penalty of buffering RAM data for each ROM access, we obtain improvements in energy consumption at the system level.
\\

\begin{figure}[t]
\centering

\includegraphics[width=0.45\textwidth]{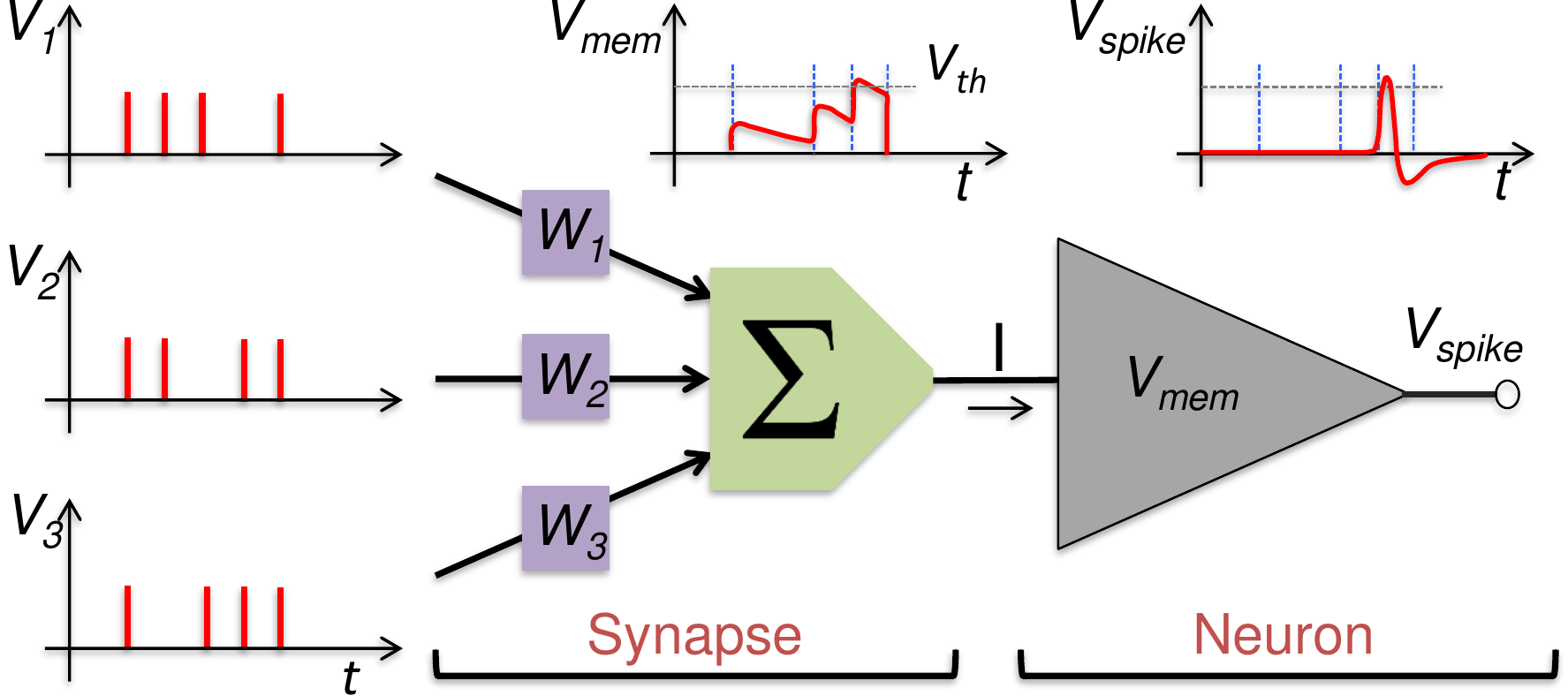}

\caption{Typical SNN dynamics. The input spikes are modulated by the synaptic weights, and the accumulated synaptic current in fed to the neuron. The neuron integrates the current and outputs a spike (fires) once its membrane potential exceeds a threshold.}
\label{fig:snn}

\end{figure}

\begin{figure}[!t]
\centering

\includegraphics[width=0.5\textwidth]{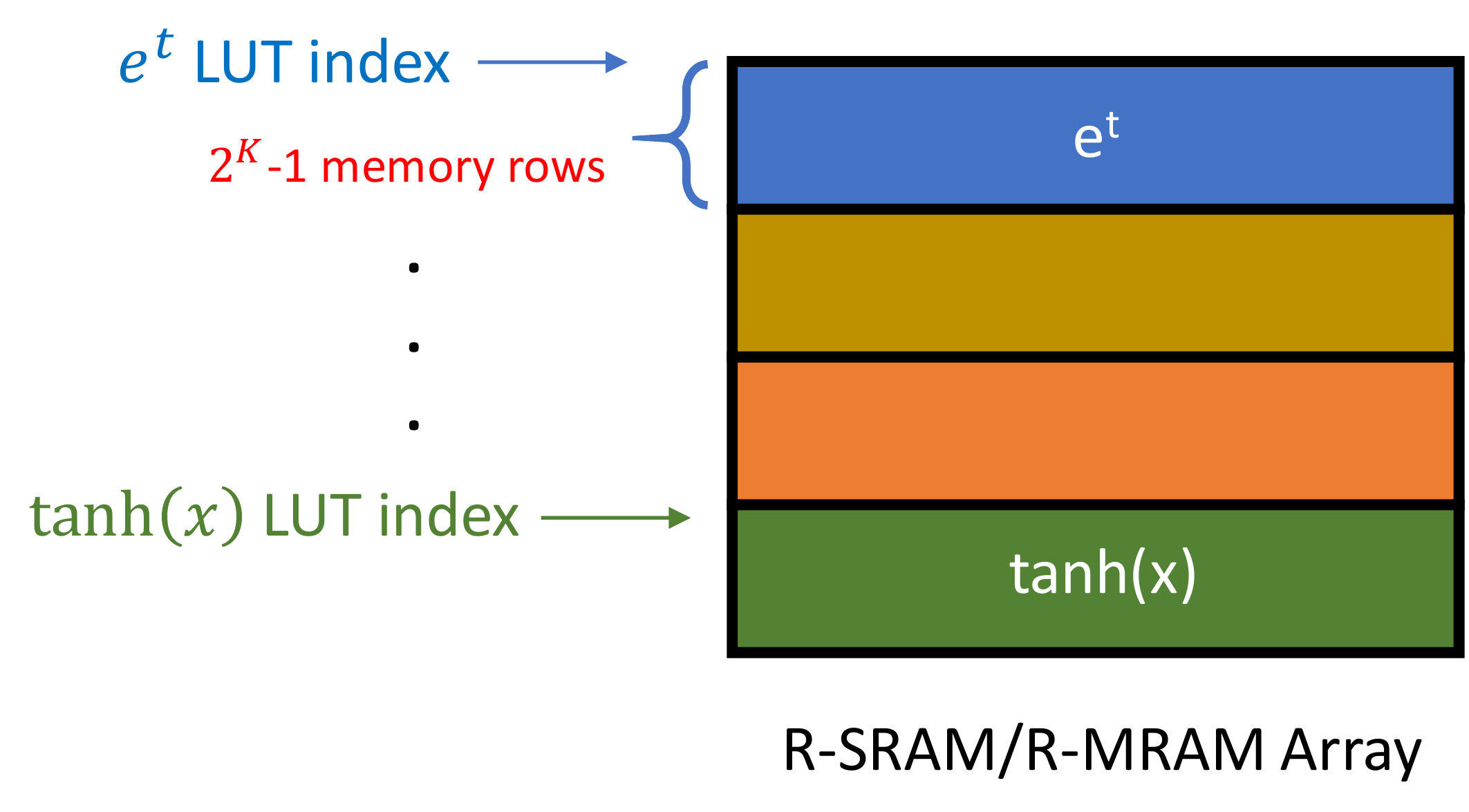}

\caption{Storage of LUTs for various functions within the same ROM-Embedded RAM array. The starting address for each type of LUT is predefined. An offset address (calculated from the input) is added to the starting address to perform the table lookup from the R-SRAM/R-MRAM. The number of memory rows required by each LUT type is predefined based on the desired precision of the transcendental function to be stored.}
\label{fig:lut}

\end{figure}

\begin{figure*}[t]
\centering

\includegraphics[width=0.85\textwidth]{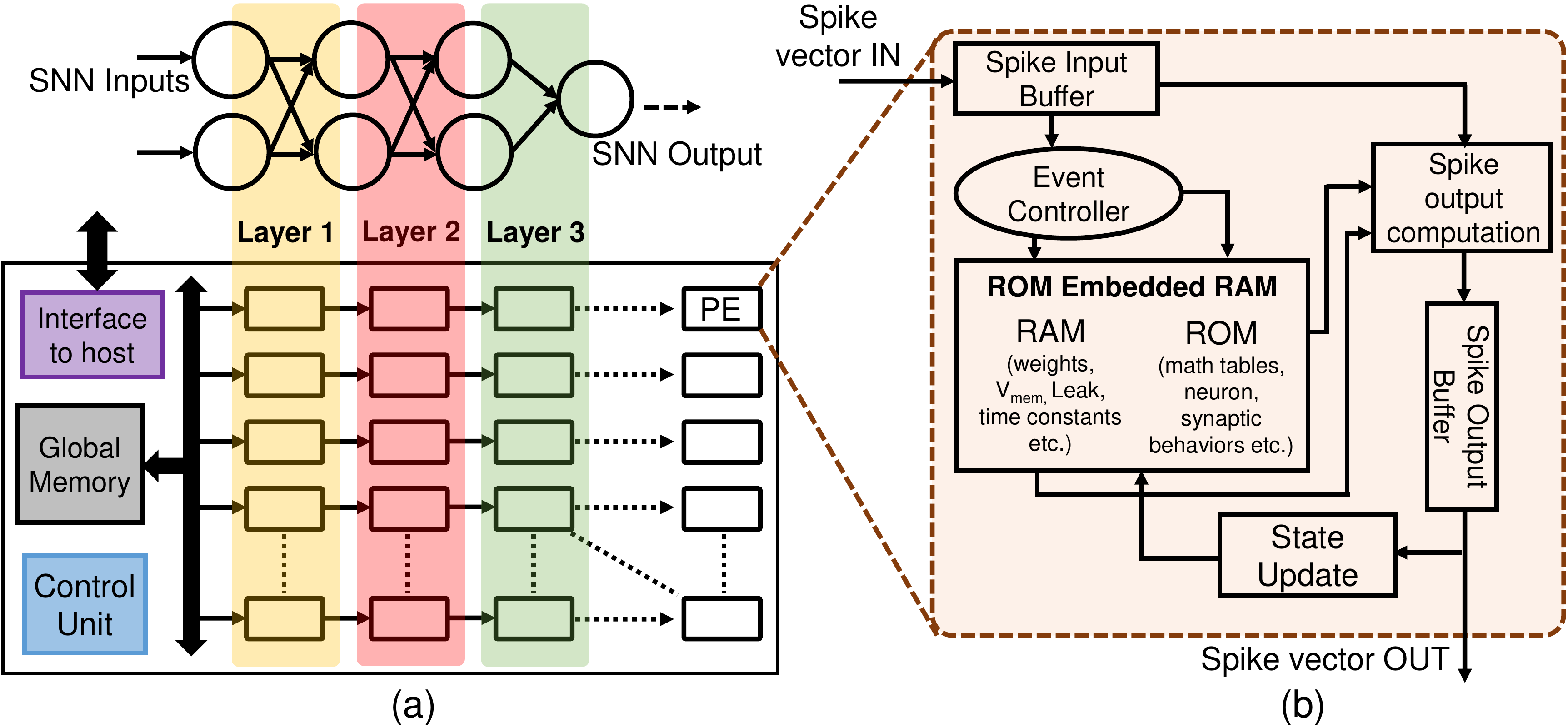}

\caption{Block level diagram of SPARE. (a) Figure shows how a deep neural network is mapped to a 2-D array of PEs connected together. The global memory stores the spiking events at every layer output, and broadcasts them to the input of next layer. (b) Figure zooms into the logical diagram of the PE. It consists of a ROM-embedded RAM to store the state variables along with LUTs of synapse, neuron and synaptic plasticity models, computation core to generate output spikes, input buffers to store incoming spike broadcast, event controller to schedule memory transactions, state updater to update the entries in the memory, and an output buffer to store the output spikes generated.}
\label{fig:arch}

\end{figure*}

\paragraph{\bf{R-MRAM}}
An R-MRAM is a memory structure made with conventional STT-MRAM array by embedding a hardware ROM. This allows it to operate in both ROM and RAM mode \cite{fong2016embedding}. As shown in Fig. \ref{fig:rmaram}, R-MRAM bit cells consist of an additional Bit Line (BL) compared to STT-MRAM. The physical connection of the bit cell (fixed during design time), stores ROM data. Bit cells connected to BL0 store ROM data `0', whereas those connected to BL1 store ROM data `1'. Every bit cell can be written/read for RAM operation by electrically connecting BL0 and BL1. However, ROM access and RAM access cannot occur simultaneously. Next we describe the R-MRAM operation for RAM mode and ROM mode of operations.

\begin{enumerate}
\item \textit{RAM mode}: During a RAM mode read operation, current from the read-bias generator flows through the pass transistors and the selected bit cell to Select Line (SL) (shown in Fig. \ref{fig:rmaram}). Consequently, a voltage appears on the positive input of the sense amplifier. The sense amplifier compares the voltage (dependent on the resistance of the selected bit cell) to a reference voltage to output a `1' or `0'. For a write operation, EnRAM is asserted to turn ON the pass transistors and the write driver drives both BLs and SL.

\item \textit{ROM mode}: For a ROM read operation, EnRAM is deassserted to turn OFF the pass transistors and the latch is turned ON. If the selected bit cell is connected to BL1 (BL0), BL1 (BL0) gets discharged and ROMOut outputs a ‘1’ (‘0’). Contrary to R-SRAM, the non-volatility of STT-MRAM prevents the RAM data to be lost in R-MRAM during a ROM read operation.
\end{enumerate}

It has been shown in prior studies that the R-MRAM design with an extra BL has no area overhead at array-level. Additionally, this doesn't impact the and performance of the memory as a ROM \cite{fong2016embedding}. Note that during ROM Mode, RAM data is not disturbed due to non-volatility of R-MRAM, thereby simplifying the ROM retrieval process. This results in higher energy benefits of using R-MRAM in SPARE, as we will show later in our simulations.

\subsection{SNN: Spiking Neural Networks}
SNN has emerged as a power-efficient choice for cognitive applications. SNNs are built using bio-plausible neurons and synapses. All information flow is converted into a train of spikes, similar to the information flow in the human brain. Refer to Fig. \ref{fig:snn}. The input spikes $V_i$ are modulated by the synapse weight $W_i$. At every time-step, $V_i$ is either `1' (spiking event) or `0' (no spike), whereas $W_i$ is a number between -1 and 1, signifying the strength of the connection. The output from all synapses is summed up and fed to the next neuron. The neuron keeps track of its membrane potential ($V_{mem}$), which gets updated based on the synaptic current. Subsequently, $V_{mem}$ accumulates/decays over several time-steps until it reaches a certain threshold $V_{th}$, when the neuron emits an output spike (`1'). This spike is then transmitted to the neurons in the next layer. Depending on the neuron model, $V_{mem}$ dynamics differ in behavior and complexity. During the training phase, the synaptic weights $W_i$ undergo changes to learn the input patterns. Many spike-based learning rules have been proposed, for example, Spike Timing Dependent Plasticity (STDP) \cite{stdp}, Long-Term Potentiation \cite{bliss1993synaptic} \textit{etc}. The basic idea is to determine the correlation between the input and output neuron spiking activities, to determine the corresponding synapse weight updates. However, once the weights are all trained, the synaptic strengths remain unchanged during the inference phase. These plasticity rules are the basis for unsupervised learning in SNNs.


\subsection{LUT based storage in R-SRAMs and R-MRAMs}

The computations required in the SNN described above rely heavily on transcendental functions and polynomial evaluations. The dynamics of $V_{mem}$, synaptic current flow, STDP learning, \textit{etc.}, all require solving differential equations with mostly exponential and higher order polynomial evaluations. The only efficient way to compute these functions in hardware is by the use of math tables or LUTs \cite{intelmathlib}. Taking an example of a typical STDP evaluation in SNNs, we show how the LUTs are structured in R-SRAM/R-MRAM.

\begin{enumerate}
\item STDP involves a synaptic weight update, based on the time difference of post- and pre- neurons ($t=t_{post}-t_{pre}$). According to this empirical rule, the change in the synaptic weight is proportional to $e^{t}$.
\item Range reduction: $t$ can have an arbitrary value. Thus, $t$ is broken into $N\frac{log2}{2^K}+r$, where K is designer's choice that determines the size of LUT, and N is $\lfloor t/\frac{log2}{2^K}\rfloor$. Thus the remainder $r$ has a confined range of $|r|\leq \frac{log2}{2^{K+1}}$. Thus, the exponential $e^t$ is reduced to $2^{N/2^K}e^r$.
\item Approximation: Due to limited range of $r$, $e^r$ can be approximated with lower order polynomials ( since $e^r=1+r+\frac{r^2}{2!}+...)$.
\item Reconstruction: To evaluate $2^{N/2^K}$, let $N=M2^K+d$, where $M=\lfloor N/2^K\rfloor$ and $d=0,1,2...2^K-1$. Thus $2^{N/2^K}=2^M2^{d/2^K}$. Using $d$ as a memory address to the R-SRAM/R-MRAM, the corresponding ROM data (LUT) is fetched, which stores $2^{d/2^K}$. The exponential reduces to $e^t=2^M\times LUT(d) \times e^r$. The multiplication by $2^M$ is a simple shift operation in hardware.
\item The exponential $e^t$ is thus used to evaluate the weight update, completing one STDP evaluation.
\end{enumerate}

Other transcendental functions and polynomials can be similarly mapped to LUTs, as described in detail in \cite{intelmathlib}. Various LUTs are stored within the same array, as shown Fig. \ref{fig:lut}. The starting address of each LUT is pre-defined and is used to perform table lookups. In the example taken above, when a `Fetch LUT' command is issued, two inputs are provided $-$ the type of LUT (exponential) and the offset (`d'). The memory address from which the lookup needs to be made is calculated by adding the offset to the LUT index corresponding to the exponential LUT.

\section{SPARE: SNN Accelerator using ROM-embedded RAMs}

\subsection{SPARE Organization}
We propose SPARE, a many-core architecture designed for efficient acceleration of SNNs. As shown in Fig. \ref{fig:arch}(a), it consists of a 2 dimensional PE-array coupled with a global memory and central control unit. A PE can perform all synapse and neuron functionalities required by different types of SNNs. This flexibility is essential as SNN computations typically differ at various levels - neurons, synapses and synaptic weight updates, depending on the application. Layers of an SNN are spatially partitioned across different PEs depending on the network size. The number of neuron state variables and synaptic weights each PE can store is limited by the memory contained within each PE. Based on the network size, number of PEs mapped to each layer are specified.

The SNN computation occurs in time-steps. At each time-step, the neuron firing data is transfered from one layer to the next. Input data spikes (for a given time-step) stored in the global memory are broadcast over the shared bus. Subsequently, the PEs mapping the first layer of the SNN start buffering the data and execute their SNN partition. Once spikes for the first layer have been transmitted, the spikes for the next layer are broadcast, and PEs mapped to second layer start their computations, and so on. 
All synaptic data is stored locally within each PE. Once the layer-1 PEs finish their execution, their output data (spikes) are written back to the global memory. Subsequently, data from all PEs is written back into the global memory, layer by layer. Consequently, this successive data transfer (neuron data) between global memory and PEs realize a time-step of SNN computation. It's worth noting that only neuron data movements occur between PEs and global memory, whereas the synapse data is locally read from the PE's RAM. This reduces the data movements in SPARE compared to a von-Neumann machine which would involve moving both neuron and synapse data between the global memory and the computation core. Additionally, this reduction is extremely significant as typical SNNs have 1000$\times$ more synapses than neurons \cite{truenorth}.

\begin{figure}[!t]
\centering

\includegraphics[width=0.5\textwidth]{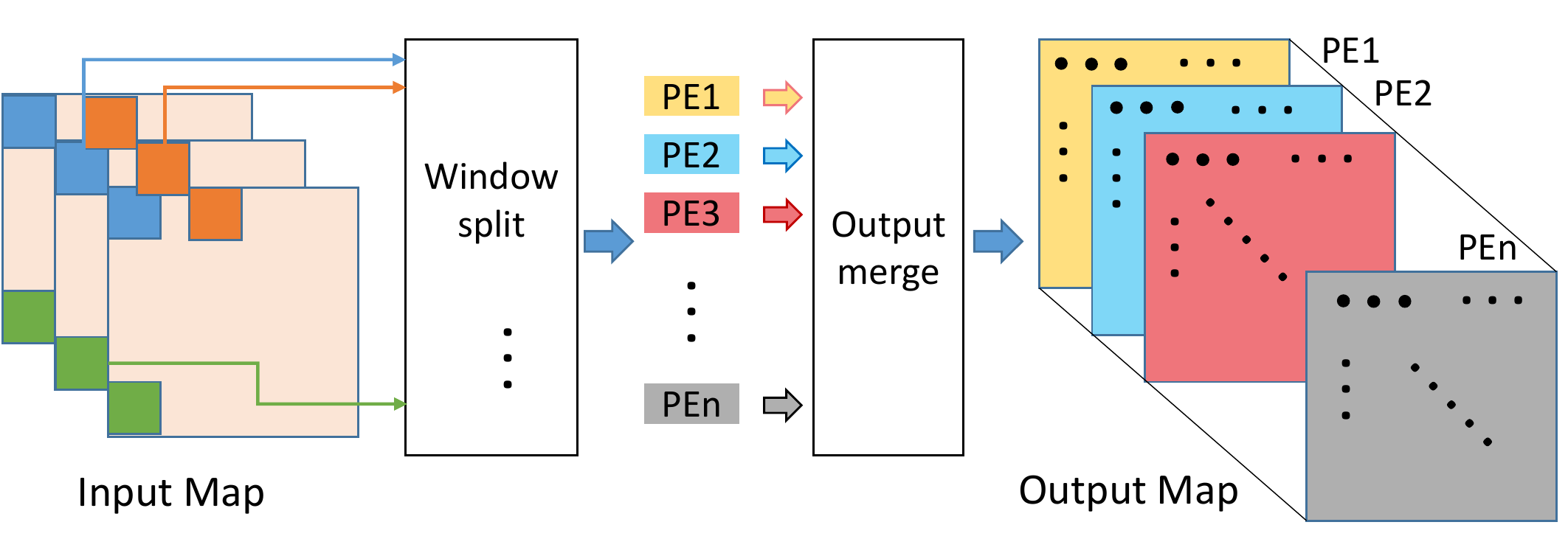}

\caption{Mapping of CNNs in SPARE: The input map is window-split based on the kernel size of the particular layer. These are then broadcast to all PEs mapped to that layer. Each PE stores different kernels, and process the data in parallel as they receive the inputs in a window split-manner. Each PE computes part of the output feature map, highlighted through color coding of PEs in figure. The output is rearranged and stored back to the global memory unit.}
\label{fig:cnn}

\end{figure}

\begin{figure}[!t]
\centering

\includegraphics[width=0.4\textwidth]{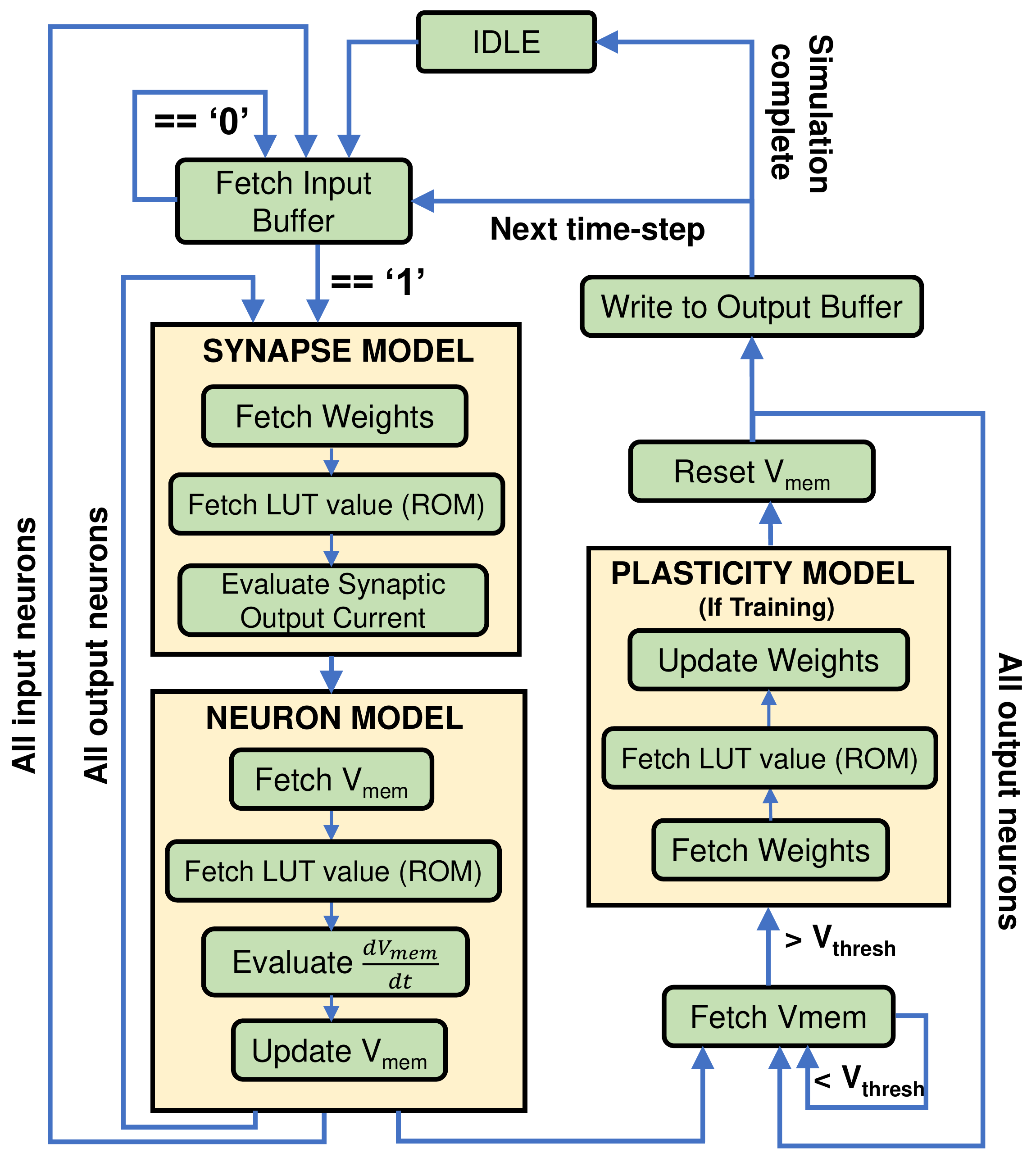}

\caption{Logical flow diagram of the event controller, describing SNN computations performed in the PE. The subsequent computation is subdivided into three main blocks. 1) Synapse model block: computes output synaptic current. 2) Neuron model block: keeps track of the membrane potential of output neurons. 3) Plasticity model block: updates synaptic weights during the training phase. This block is skipped during the inference phase.}
\label{fig:statediag}

\end{figure}

We extend this approach to map convolutional neural networks (CNNs) using SPARE. CNNs have been shown effective for image classification tasks, achieving state-of-the-art accuracies, occasionally surpassing human performance \cite{He_2016,gonature}. The standard architecture consists of alternate convolutional (c-) and spatial-pooling (s-) layers, followed by a final fully-connected (fc-) layer. Each convolutional layer hierarchically extracts complex features from the input image. This is done by using shared weight kernels that perform a convolution operation on the input image. The output of one convolutional layer becomes the input of the next. Thus, the kernels in the first convolutional layer learn low-level features, for example, edges and corners, while in deeper layers, they learn high-level features, using these low-level features as inputs. A spatial-pooling layer is added in between two convolutional layers to reduce the dimensions of the convolutional feature maps. This layer maintains the depth of the input map, however reduces the spatial dimensions. Finally, a fully-connected layer is used to determine the output class of the input image. Fig. \ref{fig:cnn} shows how the convolutional layer can be mapped to SPARE. The input map is split using a small window that strides throughout the image. The window size is governed by the kernel size of that layer. Input spikes are broadcast to the PEs in this window-split manner (instead of pixel-by-pixel manner), where each PE stores a different kernel of that layer. Thus, each PE computes part of the output feature map, which is then merged and stored in the global memory unit, as shown in the figure. Layer parameters (for example, stride, kernel size and number of output maps) are programmed into the global control unit to implement this `window-split input broadcast' and `output merge' in the global memory. Note that the s- and fc- layers can be configured as c- layers, with appropriate parameters. For s- layer, the parameters are: stride = 2, kernel size = $2x2$, number of output maps = number of input maps. Whereas for an fc- layer, stride = 0, kernel size = input feature size, number of output maps = number of output neurons. Thus, the proposed architecture is a generalized programmable architecture that maps convolutional, spatial pooling as well as fully-connected layers.

\begin{figure}[!t]
\centering

\includegraphics[width=0.4\textwidth]{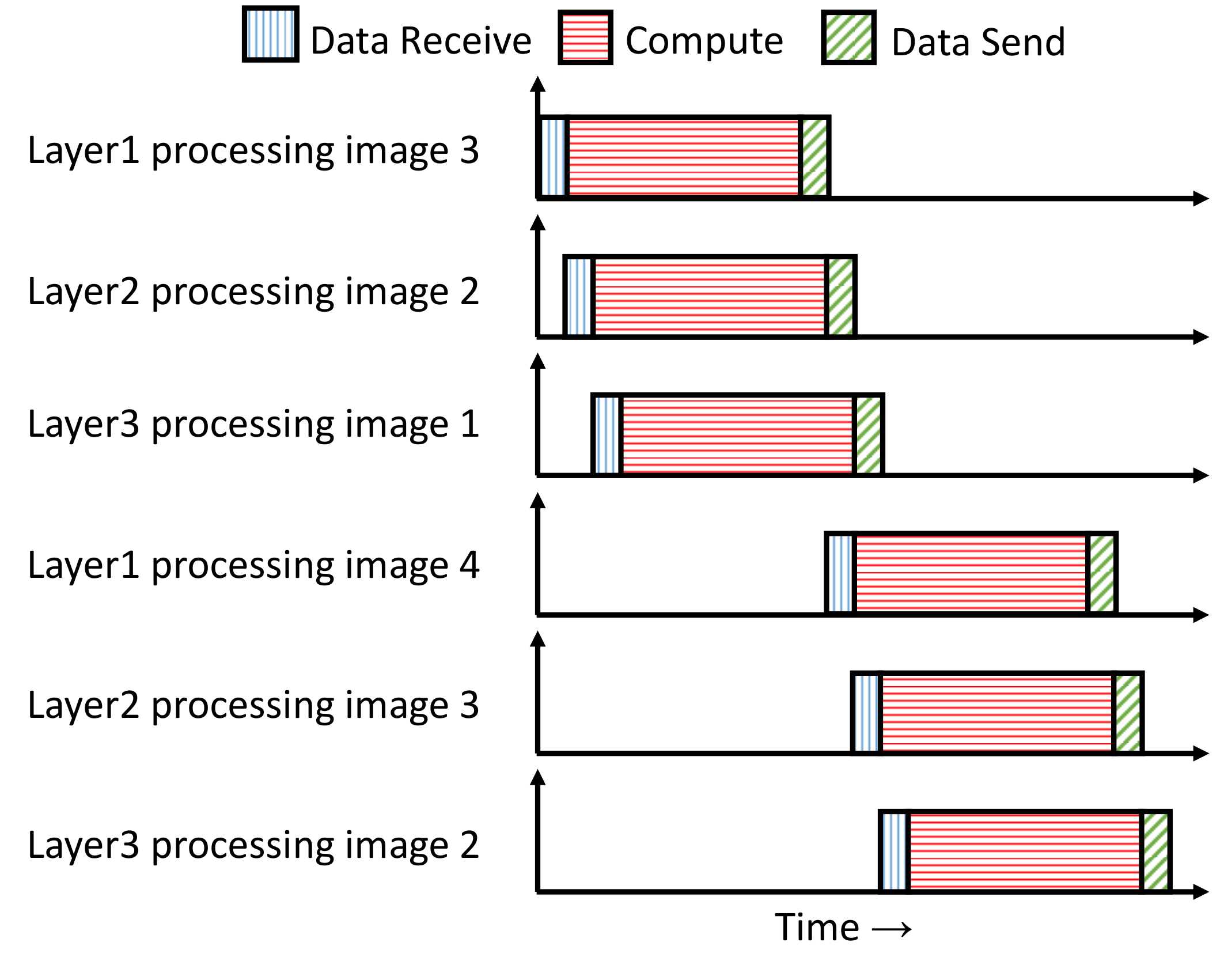}

\caption{Timing diagram illustrating the inter-layer pipelining in SPARE. As soon as the PEs receive and buffer the input data, they start processing. Meanwhile, data for PEs mapped to subsequent layers is transmitted. Since the data transfer time is small compared to the computation time within the PE, all PEs process data in parallel.
}
\label{fig:pipe}

\end{figure}

\subsection{Inter-layer pipelining}
SPARE enables a pipelined execution of layers in an SNN to exploit the available inter-layer data parallelism. Layers of SNN are mapped across the 2-dimensional PE array. Hence, while layer-2 PEs are computing the n\textsuperscript{th} input image, layer-1 PEs compute the (n+1)\textsuperscript{th} input image and so on. Data communication between layers of SNN are achieved by scatter and gather operations initiated by the SPARE control unit (see Fig. \ref{fig:arch}(a)) to move data between global memory and PE-array. SPARE control unit stores the mapping information between SNN layers and PEs. A gather operation for a layer collects the output data computed by the PEs mapped to the specific layer and stores it in the global memory. Scatter operation for a layer sends the input data to the required PEs. It is important to note that data communication in SNNs is of feed-forward nature where PEs mapped to layer-n will only send data to PEs mapped to the subsequent layer-n+1 and so on. Hence, we do not support a dedicated on-chip network for all PE-to-PE communication due to the associated area and power overheads. Our ``in-memory'' nature of computing results into PEs spending more time in computation (within PE) rather than sending and reeving data from global memory. Hence, our inter-layer communication based on a shared resource (global memory) doesn't lead to performance issues due to the natural pipelining obtained as shown in Fig \ref{fig:pipe}. 

\begin{figure}[!t]
\centering
\includegraphics[width=0.45\textwidth]{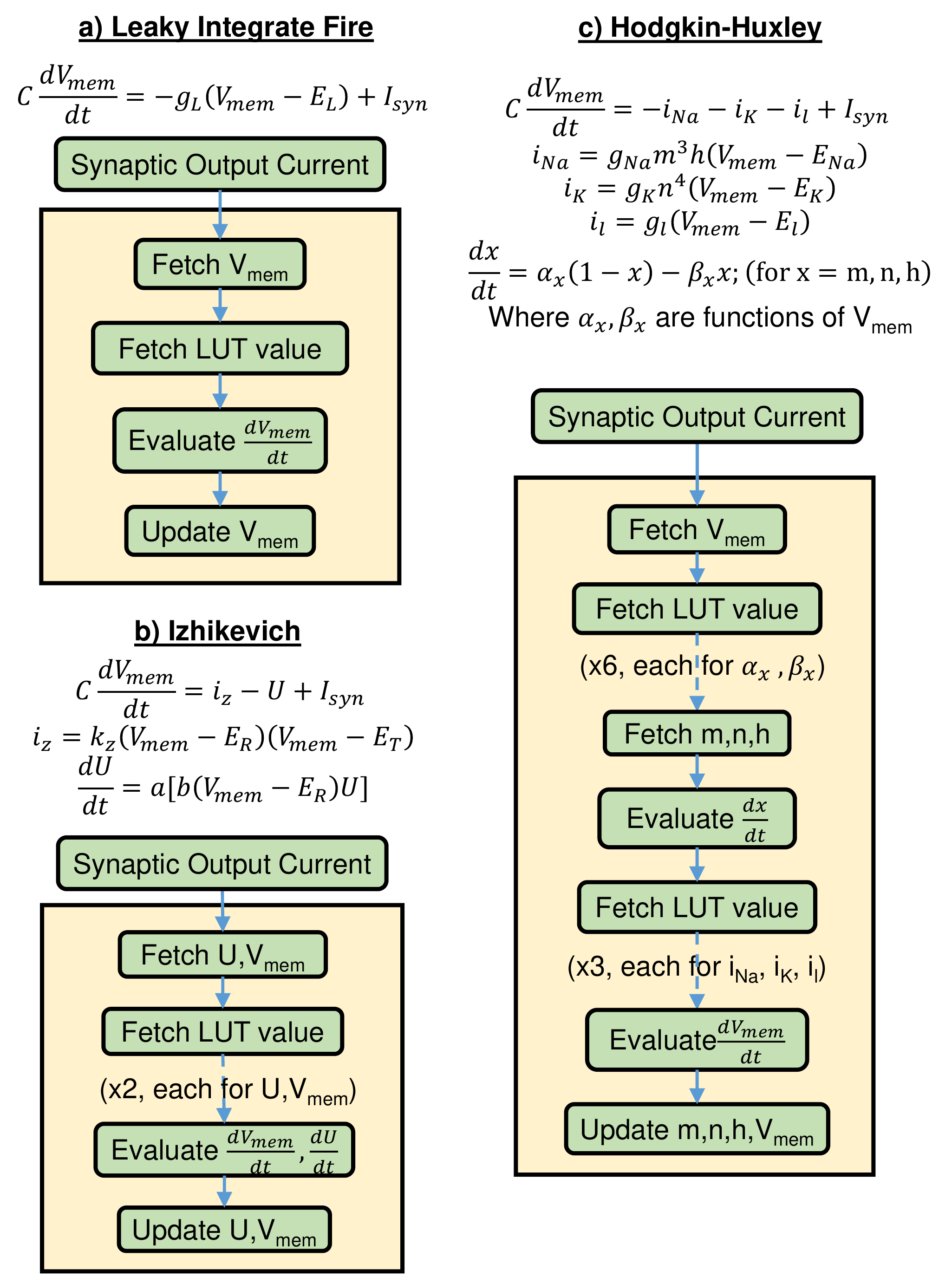}
\caption{Differential equations describing the dynamics of neurons and an LUT based approach to implement them in SPARE. (a) Leaky-integrate-fire neuron (b) Izhikevich neuron (c) Hodgkin-Huxley neuron}
\label{fig:neuronmodels}
\end{figure}

\begin{figure*}[!t]
\centering
\includegraphics[width=0.9\textwidth]{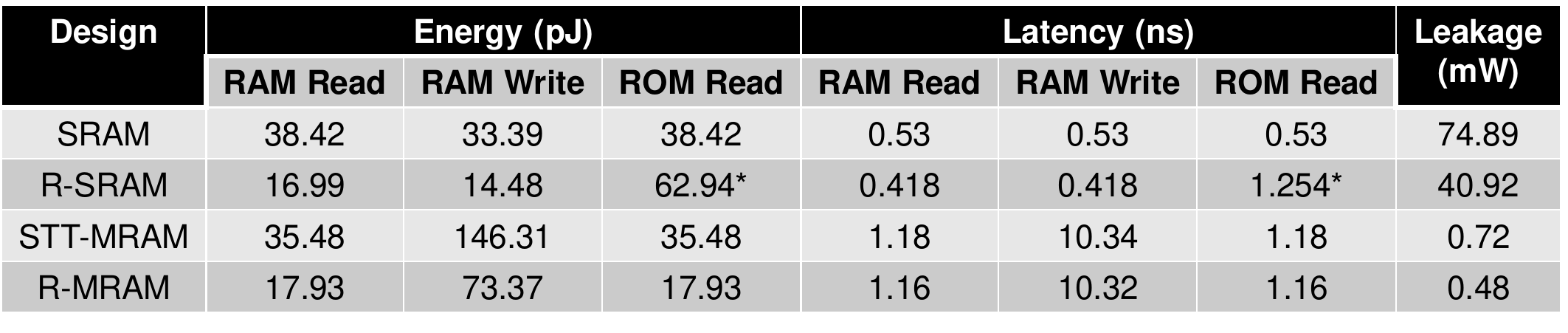}
\caption{Energy and latency for read-write accesses from all designs considered in this work $-$ SRAM, R-SRAM, STT-MRAM, and R-MRAM. (* ROM Read for R-SRAM includes additional overhead of buffering RAM, retrieving ROM data and storing back the buffered RAM data, as described in Section \ref{sec:roeam}).}
\label{fig:numbers}
\end{figure*}

\begin{figure}[t]
\centering
\includegraphics[width=0.4\textwidth]{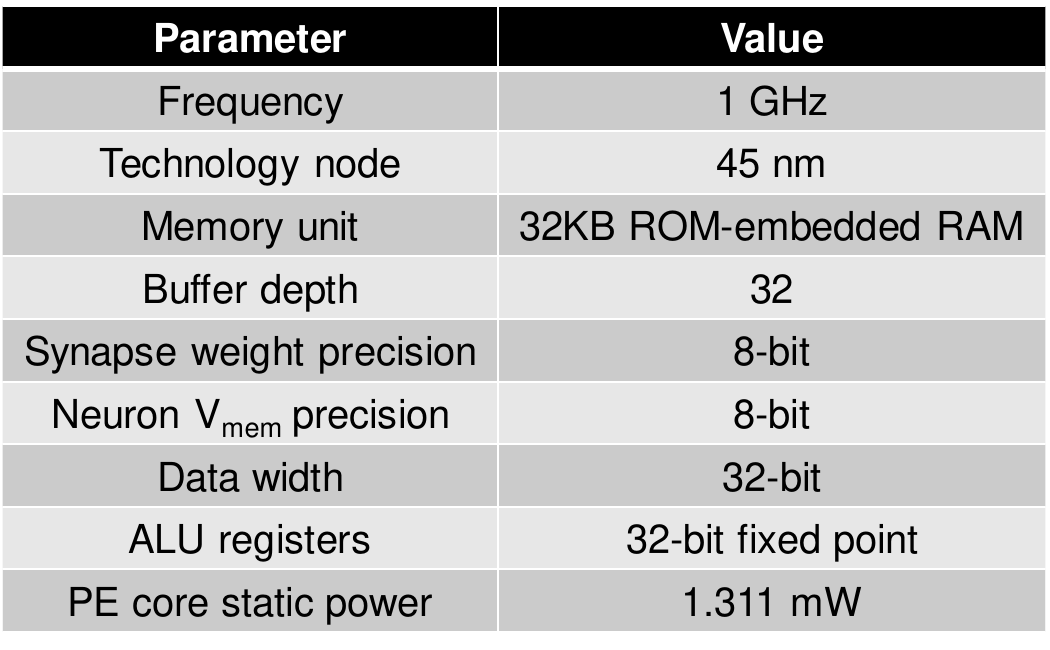}
\caption{uArchitecture design parameters used for simulations. }
\label{fig:table1}
\end{figure}

\subsection{Processing Element (PE)}
\label{sec:pe}
As shown in Fig. \ref{fig:arch}(b), PE contains a computing core to perform SNN computations and a memory unit to store the neuron-synapse models, state variables and LUTs. The memory unit (RAM and ROM) and the computation core within the PE localize most of the data movements required for computing the output neurons (mapped to the PE), thereby enabling ``in-memory processing''. While RAM houses all the synaptic weights and state variables required for the output neuron computations, the ROM stores the LUTs required for modeling synapse, neuron and synaptic weight update computations. Consequently, the higher storage density enabled by ROM-embedded RAMs (smaller memory size) and the resulting reduction in data movements increases the computation efficiency and reduces overall energy consumption. The computational flow in a PE and a step-by-step procedure for typical SNN computation is illustrated in Fig. \ref{fig:statediag}. It consists of three main blocks: 1) Synapse model, 2) Neuron model and 3) Plasticity model. The event controller checks the head of the input spike buffer, and both the Synapse and Neuron blocks are skipped if the input is `0', thereby leveraging the benefits of event-driven computing in SNNs to achieve energy-efficiency. Similarly, the Plasticity block is skipped if the $V_{mem}$ is less than the threshold (no synaptic weight update). PEs are modeled as extended finite-state machines. As soon as the PE receives the broadcast of spikes corresponding to its layer tag, it starts computing. Thus, effectively all PEs run in parallel, exploiting data-parallelism and inter-layer pipelining. Since the input spikes are broadcast to all PEs, each PE performs the SNN computation corresponding to the neuron and synapses it is mapped to. Since SPARE localizes data-movement through in-memory computing, the same memory storage unit also contains the LUTs used in SNN computations allowing a simple and compact PE design.


\subsection{Modeling complex neuro-synaptic functionality}
\label{sec:hh}
Most neuron-synapse models have complicated differential equations, and heavily use higher order polynomials and transcendental functions. This makes them highly suitable for an LUT based storage in ROM-embedded RAMs. Thus, our PE incorporates any model needed by the SNN application without much area overhead. To illustrate this, dynamics of three different neuron models - LIF \cite{lif}, Izhikevich \cite{izhi}, and Hodgkin-Huxley (HH) \cite{hh} are shown in Fig. \ref{fig:neuronmodels}. Each model can be implemented in SPARE, by modifying the `neuron model' block in the state diagram (see Fig. \ref{fig:statediag}), with corresponding alterations. As shown in Fig. \ref{fig:neuronmodels}(c), the HH model is described by 7 differential equations, with 4 state variables - $V_{mem}, m,n$ and $h$. Firstly, we need 4 RAM fetches to read the state variables. To update $m,n,$ and $h$, we need a total of 6 ROM fetches, each for $\alpha _{m,n,h},\beta_{m,n,h}$. Next, we need to evaluate higher order polynomials (also stored in LUTs) in order to calculate $i_{Na,K,l}$. Thus, a total of 9 ROM fetches, 4 RAM fetches and 4 RAM updates per spike per time-step are required for HH model, in contrast to 1 ROM fetch, 1 RAM fetch and 1 RAM update in case of a simple LIF model. However, the overall data flow diagram remains unchanged. A similar approach can be used to implement various synapse and plasticity models, by modifying the synapse and plasticity blocks, respectively. As the models become more complex, more LUTs are required to store multiple polynomial functions and math tables. Moreover, the number of ROM and RAM fetches also increase. SPARE addresses both these issues since the ROM-embedded RAM primitive allows \textit{extra ROM} for LUT storage, thereby allowing a compact memory unit. Data-localization in SPARE along with the compact memory storage unit enables a lower energy/latency per ROM/RAM access.

\begin{figure*}[t]
\centering
\includegraphics[width=\textwidth]{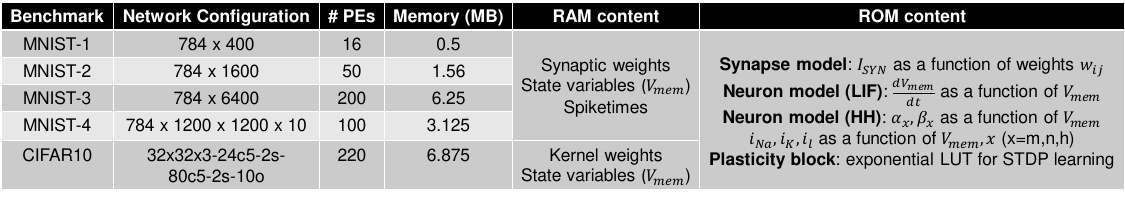}
\caption{SNN benchmarks used in SPARE evaluation \cite{diehl2015fast,Diehl_2015}. The figure tabulates the number of PEs required, memory requirement, and the RAM/ROM content for each benchmark and neuron model.}
\label{fig:table2}
\end{figure*}

\section{Experimental Methodology}
PE was modeled at the Register Transfer Level (RTL) in Verilog and synthesized to IBM 45nm technology library using Synopsys Design Compiler to estimate the power and area consumptions. R-SRAM and R-MRAM (memory units) were modeled using Cacti \cite{cacti} and NVSim \cite{nvsim}, respectively, for the corresponding RAM sizes at 45nm technology node. Subsequently, we account for the modified ROM access cycles and peripheral circuits described in Sec. \ref{sec:roeam}. Fig. \ref{fig:numbers} summarizes the RAM/ROM read-write energy and latency obtained from simulations for SRAM, R-SRAM, STT-MRAM and R-MRAMs. Cycle-accurate RTL simulations were performed to get estimates of memory (RAM, ROM) access traces and subsequently, the overall energy consumption per classification. Fig. \ref{fig:table1} summarizes various $\mu$-architecture parameters used for the simulations.


We analyze the energy, performance and area benefits of SPARE on MNIST dataset \cite{mnist} and CIFAR-10 dataset \cite{cifar10}. For an apples-to-apples comparison, we use a similar architecture built with PEs comprised of typical RAM and STT-MRAM as our baselines (without ROM-Embedded RAM capability). Additionally, to demonstrate system scalability, we benchmark SPARE with various network sizes of different scales, varying from 1184 to 36602 neurons. Fig. \ref{fig:table2} tabulates the benchmarks chosen \cite{Diehl_2015,diehl2015fast,bing}, and the number of PEs required in each case. Note that benchmarks `MNIST-1,2,3' are typical two-layer SNNs, that can be trained using STDP learning \cite{Diehl_2015}. The input layer has 784 neurons, each corresponding to an input pixel in the image. The output layer has 400, 1600 and 6400 neurons for benchmark `MNIST-1,2 and 3' respectively. For deep spiking networks beyond two-layers, there hasn't been any successful attempt to generalize a training algorithm in the spiking domain. However, \cite{diehl2015fast,bing} show that off-line training of the network using DNN techniques (standard back-propagation algorithm) and converting the trained network to an SNN does not incur significant performance degradation. Thus, to evaluate SPARE on deep networks, we use benchmarks `MNIST-4' and `CIFAR10', in the inference phase. `MNIST-4' is a deep multi-layered, fully connected SNN converted from a trained DNN \cite{diehl2015fast}. It consists of an input layer of 784 neurons, followed by two hidden layers with 1200 neurons each, and finally an output layer of 10 neurons. Benchmark `CIFAR10', on the other hand, is a deep convolutional neural network converted from a trained CNN (32x32x3-24c5-2s-80c5-2s-10o). The CNN has two convolution (c-) and two spatial-pool (s-) layers arranged alternately, followed by a fully-connected (fc-) output layer. The dimension of input image is 32x32x3. The first c- layer consists of 24 kernels of size 5x5x3. The following s- layer has kernels with size 2x2. The second c- layer has 80 kernels of size 5x5x24, followed by another s- layer with kernel size 2x2. The final layer has 10 neurons, fully connected to the previous layer. In all our simulations, we use the LIF neuron model along with the exponential STDP based plasticity for the training phase.

\begin{figure*}[t]
\centering
\includegraphics[width=\textwidth]{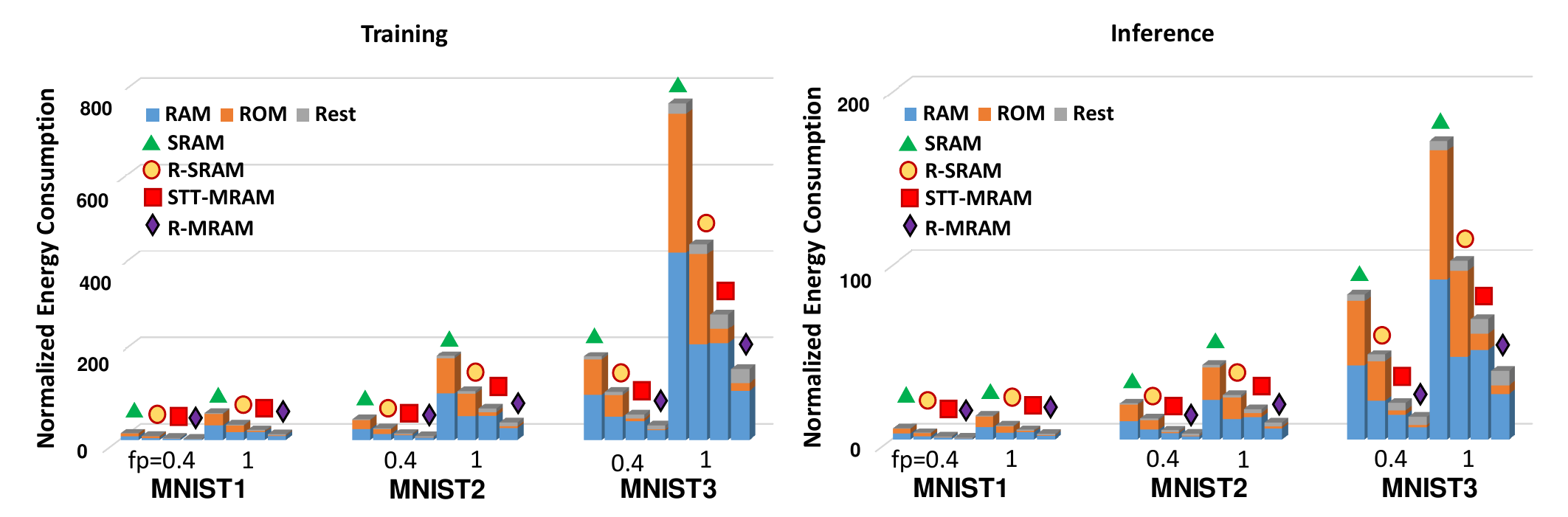}
\caption{Normalized energy consumption for a) Training phase and b) Inference phase, for benchmarks `MNIST1-3'. The simulations are performed for max firing rate $fp=0.4$ and $1$. The energy bars are further split into RAM (read/write energy + leakage), ROM (read energy + leakage) and Rest (core energy). The energy values are normalized to the common base reference.}
\label{fig:energy123}
\end{figure*}

\begin{figure}[t]
\centering
\includegraphics[width=0.5\textwidth]{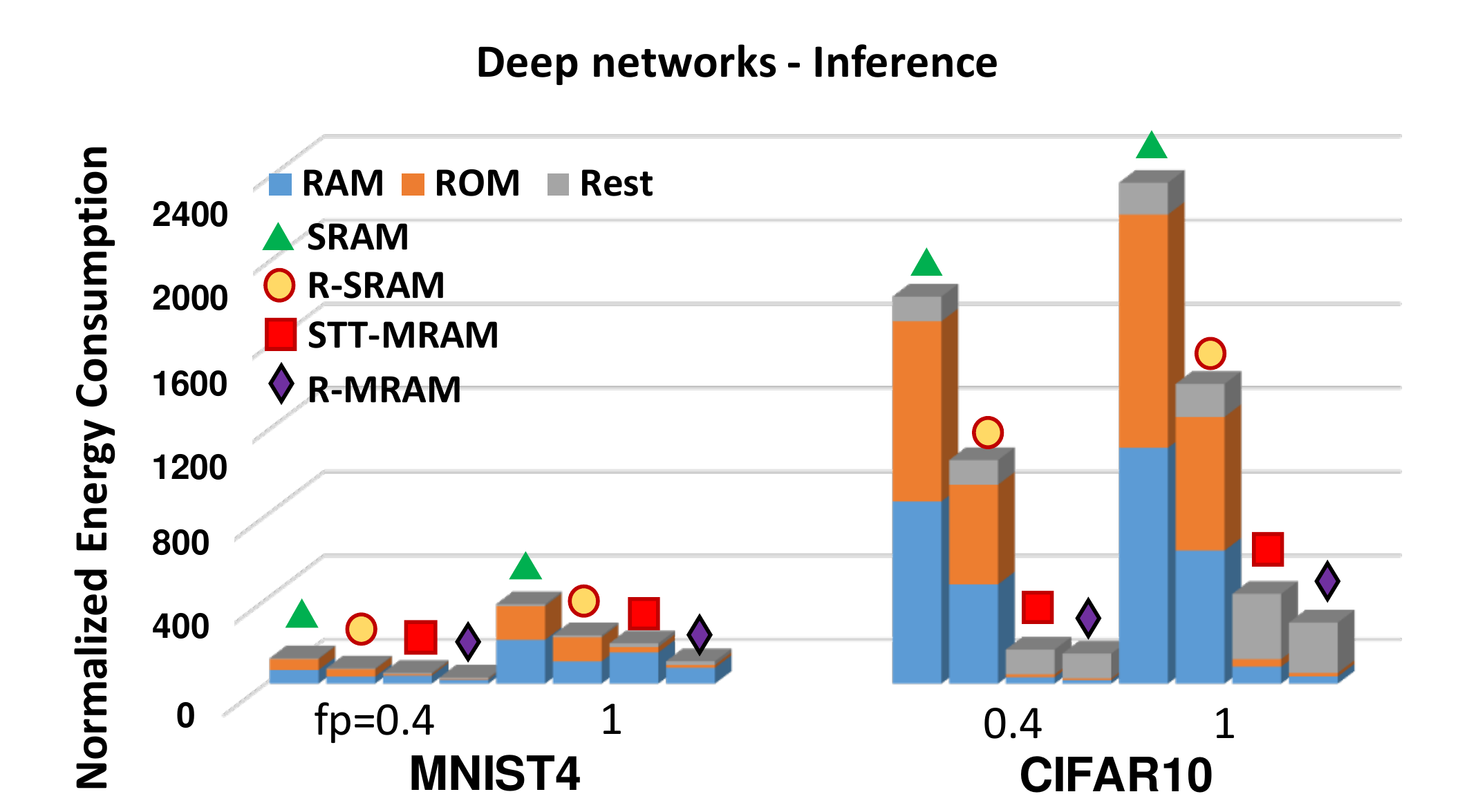}
\caption{Normalized energy consumption for benchmarks `MNIST4' and `CIFAR10'. The simulations are performed for max firing rate $fp=0.4$ and $1$. The energy bars are further split into RAM (read/write energy + leakage), ROM (read energy + leakage) and Rest (core energy). The energy values are normalized to the common base reference.}
\label{fig:energy45}
\end{figure}

Input spike trains were generated from the input image pixels based on the rate-coding approach used in \cite{brian}. Each image is split-up into several time-steps, each conveying the input firing activity. We analyze the benefits of SPARE towards leveraging SNN data sparsity (event-drivenness) by analyzing each SNN on different input maximum firing rates, $fp=0.4$ and $fp=1$ \cite{bing}. Kindly note that in our work, we use the SNN size and dataset statistics only for exploring system scalability and benefits of in-memory computing for training and testing SNNs. Mapping of these networks to the proposed architecture doesn't lead to any degradation in the classification accuracy. Readers are referred to \cite{Diehl_2015,diehl2015fast,bing} to explore the classification accuracy achieved in the benchmarks used.




\section{Results}

\subsection{Energy}
\label{sec:energy}
\begin{figure*}[t]
\centering
\includegraphics[width=\textwidth]{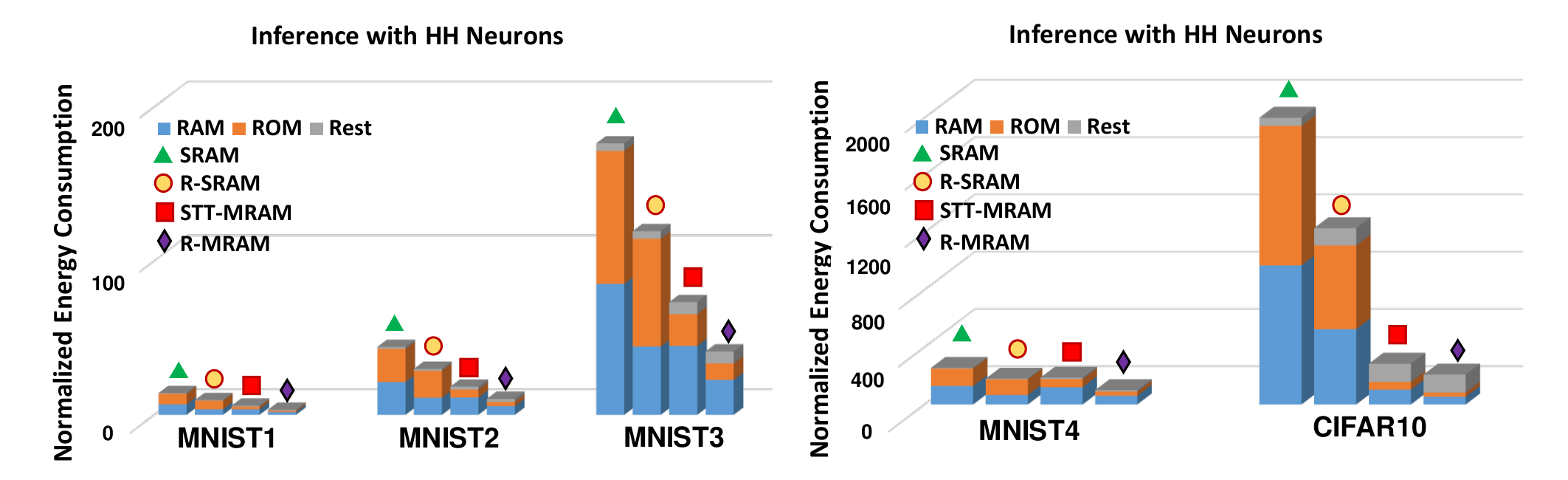}
\caption{Normalized energy consumption for using Hodgkin-Huxley neuron models on SNN benchmarks. The simulations are performed for max firing rate $fp=0.4$. The energy bars are further split into RAM (read/write energy + leakage), ROM (read energy + leakage) and Rest (core energy). The energy values are normalized to the common base reference.}
\label{fig:energyHH}
\end{figure*}

A common base reference was used to normalize all energy numbers obtained through simulations such that the minimum energy consumption bar (Inference of MNIST-1 for R-MRAM with fp=0.4) represents 1. All other energy bars in Fig. \ref{fig:energy123}, Fig. \ref{fig:energy45}, and Fig. \ref{fig:energyHH} are normalized to this value.
Fig. \ref{fig:energy123} shows the energy consumption for benchmarks `MNIST-1,2,3', both for training and inference phases. Each bar shows the total energy, which is further split into three sub-components 1. RAM (access + leakage) 2. ROM (access + leakage) and 3. Rest (Core - buffer, control, compute). The following observations can be drawn from Fig. \ref{fig:energy123}. 1) It can be seen that an increase in the maximum firing rate (fp) results in increased overall energy consumption across all datasets. This is because a higher firing rate results in increased number of spikes. Consequently, this increases the number of RAM/ROM accesses, thereby decreasing the benefits from event-driven computing in SPARE. This also increases the overall computations as more synapses will be accumulated over the output neurons. This underscores the effectiveness of SPARE in drawing benefits from the data sparsity in SNNs. 2) The total energy consumption in the inference phase is lower compared to the training phase because the Plasticity block (refer Fig. \ref{fig:statediag}) is skipped during the inference phase, as described in Sec \ref{sec:pe}. 3) The energy consumption with STT-MRAM technology is more than $\sim 2\times$ less, compared to CMOS based memory technology. 

This was expected since STT-MRAM is a NVM, thus, leakage due to memory is close to 0. Although writing into STT-MRAM is expensive compared to CMOS, the near-zero leakage is a dominant factor in reducing the energy consumption. 4) Using R-SRAM and R-MRAM as the memory units in the PE, we obtain, $1.71\times$, and $1.76\times$ reduction in energy consumption on an average, compared to CMOS SRAM and STT-MRAM, respectively. This is a direct consequence of increased storage density (or, smaller area for iso-bytes) provided by ROM-embedded RAMs. A smaller memory reduces the access energy and the leakage, thereby leading to energy benefits. However, note that for iso-area, higher storage density (through ROM-embedded RAMs) allows bigger on-PE storage, eliminating data movements required from external memory (in case of typical RAM). This leads to energy benefits. Note that we have used iso-storage PEs in our simulations to evaluate the energy benefits. 5) The energy improvement in STT-MRAM technology is greater compared to CMOS due to a simpler ROM retrieval process in R-MRAMs compared to R-SRAM (refer Sec. \ref{sec:roeam}). R-SRAMs require additional steps in buffering the RAM data, for each ROM access, which is not required in R-MRAMs.

Moving to deeper networks, Fig. \ref{fig:energy45} shows the energy consumption for deep networks, illustrating the scalability of SPARE towards executing SNN workloads. A few additional observations can be inferred: 1) Benchmark `MNIST4' being a deeper extension of `MNIST1-3', obtains similar improvements of $1.65\times$, and $1.77\times$ reduction in energy consumption for CMOS and STT-MRAM technologies, respectively. 2) For a deep convolutional network (`CIFAR10'), the improvements are $1.70\times$ and $1.31\times$, for CMOS and STT-MRAM, respectively. CNNs are more compute-intensive compared to memory-intensive fully connected networks \cite{cnncomputeintensive}. Thus, more energy is spent in computations, compared to the memory transactions. Thus, the energy consumed by the core and the memory leakage energy are significant. For the CMOS case in benchmark `CIFAR10', the memory leakage overwhelms the core energy consumption (see Fig. \ref{fig:energy45}), whereas in STT-MRAM, the core energy consumption overwhelms the memory energy (no leakage!). Due to this reason, the improvement of using R-MRAMs is suppressed in CNNs. Comparing only the memory energy consumption (RAM+ROM), we still obtain $\sim 2\times$ improvement for R-MRAMs, however, the core energy being dominant reduces overall benefits. Note that using the STT-MRAM technology itself decreases the energy consumption by an order of magnitude compared to CMOS. Thus, we conclude that using R-SRAMs over typical SRAMs as compute units lead to $\sim 1.7\times$ improvement in energy for both fully-connected networks and convolutional networks. Whereas, using R-MRAMs over typical STT-MRAMs lead to $\sim 1.75\times$ improvement for fully-connected networks, and $\sim 1.3\times$ for CNNs.

\begin{figure}[t]
\centering
\includegraphics[width=0.5\textwidth]{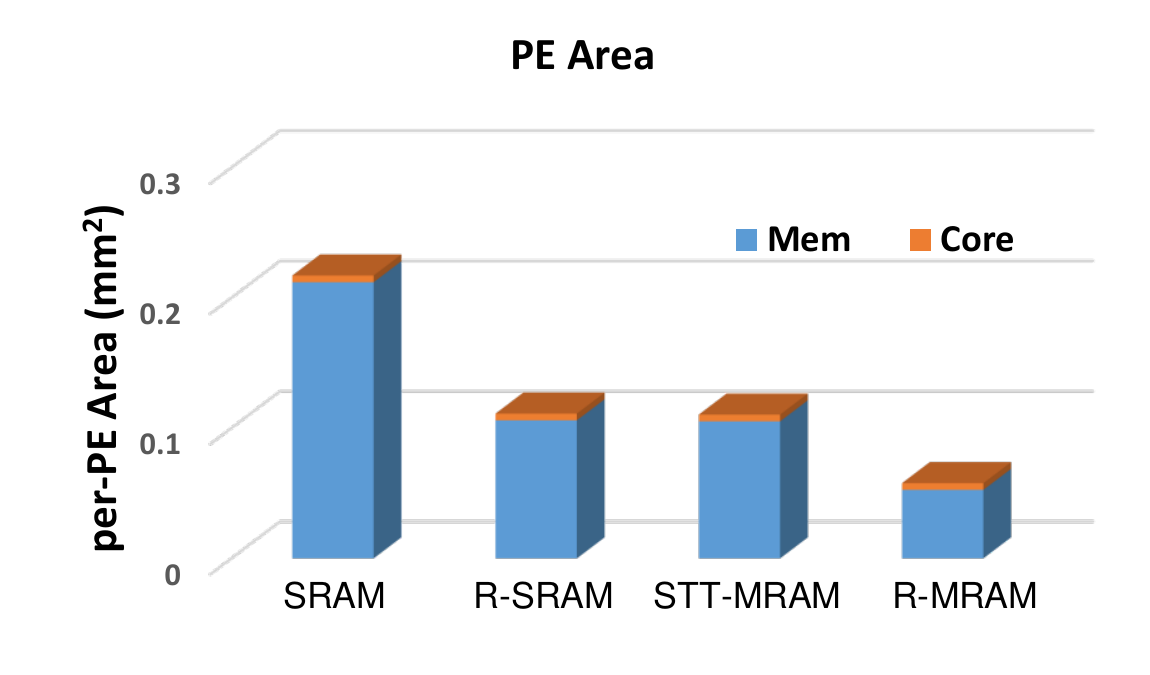}
\caption{per-PE area with SRAM, R-SRAM, STT-MRAM and R-MRAM as memory units (for iso-storage).}
\label{fig:area}
\end{figure}

\subsection{Area}
By using R-SRAMs and R-MRAMs in PEs, $1.95\times$ and $1.91\times$ area benefits are achieved on a per-PE basis, for R-SRAM and R-MRAM, respectively, shown in Fig. \ref{fig:area}. This is because ROM-embedded RAM effectively provides extra ROM with no area overhead. Moreover, the PE area is dominated by the memory unit, since the core (buffers, controller and computation core) consumes a small portion of the total area.  The area consumption of the shared bus and global memory are insignificant with respect to the total PE area (hundreds of PEs used in benchmarks - see Fig. \ref{fig:table2}). This translates to SPARE being more area-efficient compared to a normal-RAM based system.

\subsection{Performance}

In the previous section, we observed a reduction in PE area by a factor of $1.91-1.95\times$ by using R-SRAMs/R-MRAMs, due to higher storage density provided by ROM-embedded RAMs. For a given chip area (iso-area), we can fit about twice as many PEs that use R-SRAM and R-MRAM compared to typical SRAMs and STT-MRAMs, respectively, by using smaller memory sizes. Computations (neurons in a layer) can be split between more PEs, translating to $1.91-1.95\times$ performance benefits. Note that we assume a ROM:RAM ratio of 1:1. This is reasonable for SNN computations due to extensive LUT demands arising from various math function requirements. However, if the designer wishes to decrease the ratio (at the cost of lower precision of LUTs and lower flexibility with respect to neuro-synaptic functionalities), the performance improvement would be smaller, as the ratio decreases.






\subsection{Complex neuro-synaptic models}

We expect to achieve higher benefits in mapping more complicated neuro-synaptic models, due to increased LUT storage demands and ROM accesses per classification. In literature, usage of complicated models, such as the Hodgkin-Huxley (HH) and Izhikevich neuron models, is limited to biological experiments, and no references report a decent classification accuracy in using such models in SNN classification tasks. However, in order to evaluate SPARE for more complex models, we estimate the energy benefits of using HH neuron model for the same benchmarks used before. Note that the level of complexity of the differential equations increases from LIF to Izhikevich to HH, as described earlier in Section \ref{sec:hh}. For LIF, 1 ROM fetch, 1 RAM write, and 1 RAM read is required per computation. For Izhikevich, 2 ROM fetches, 2 RAM writes, and 2 RAM reads are required. While for HH, 9 ROM fetches, 4 RAM writes, and 4 RAM reads are needed. Thus, it is trivial that the energy consumption would increase as we go from LIF to Izhikevich to HH. To avoid clutter, we only compare the two extreme cases (LIF and HH), to evaluate SPARE with complex neuron models.
Fig. \ref{fig:energyHH} shows the normalized energy consumption for the inference phase for MNIST and CIFAR10 datasets, using the HH neuron model at max firing rate $fp=0.4$. Note that these are only projected values showing the energy profiles in using HH neurons for SNN workloads. The following observations can be inferred: 1) The energy consumption is higher, as compared to the LIF neuron implementation, throughout all datasets. Moreover, energy spent in ROM accesses is higher than RAM accesses. This is a direct consequence of additional RAM and ROM fetches (9 ROM fetches, 4 RAM fetches, 4 RAM updates per spike per timestep) involved in solving complex differential equations for HH neurons. 2) For fully connected networks, we obtain $1.45\times$ and $1.84\times$ reduction in energy consumption on an average, for R-SRAMs and R-MRAMs compared to CMOS SRAM and STT-MRAM, respectively. While for convolutional networks, we obtain $1.67\times$ and $1.4\times$ reduction. Note that the corresponding improvements in energy are higher for R-MRAM technology, but lower for the R-SRAM technology, compared to the LIF neuron case (Sec. \ref{sec:energy}). This is due to the fact that R-SRAMs have additional overhead in ROM retrieval process, as described earlier. Since HH neurons involve lots of ROM accesses, this overhead leads to a reduction in energy improvements. While for R-MRAMs, since the overhead is minimal, increased ROM accesses leads to higher energy benefits.

\section{Conclusion}


In this paper, we presented SPARE, an architecture utilizing the `in-memory processing' abilities of ROM-embedded RAM to enable efficient acceleration of SNNs. Each processing unit in SPARE does event-driven processing in order to leverage the benefits from input data sparsity in SNNs.  We analyzed trade-offs of using CMOS based R-SRAMs and STT-MRAM based R-MRAMs as memory units in SPARE for different types of networks. Our experiments on various SNN benchmarks for image classification applications reveal that R-MRAMs are suitable for mapping fully-connected networks compared to typical STT-MRAM arrays with $\sim 1.75\times$ lower energy, while R-SRAMs are suitable for mapping CNNs compared to typical SRAM arrays with $\sim 1.7\times$ lower energy. R-SRAM and R-MRAM achieve $\sim1.9\times$ reduction in area for iso-storage. Moreover, for iso-area, R-SRAMs and R-MRAMs can achieve $\sim1.9\times$ improvement in performance, given required data parallelism (neurons in a layer) is available. SPARE also provides the necessary programmability to execute a variety of synapse, neuron and plasticity models thereby, enabling designers to deploy SNNs based on the application requirements. SPARE thus underscores the applicability of ROM-embedded RAM based in-memory hardware primitives in efficient cognitive computing.






\bibliography{refbib}
\bibliographystyle{ieeetr}

\begin{IEEEbiography}[{\includegraphics[width=1in,height=1.25in,clip,keepaspectratio]{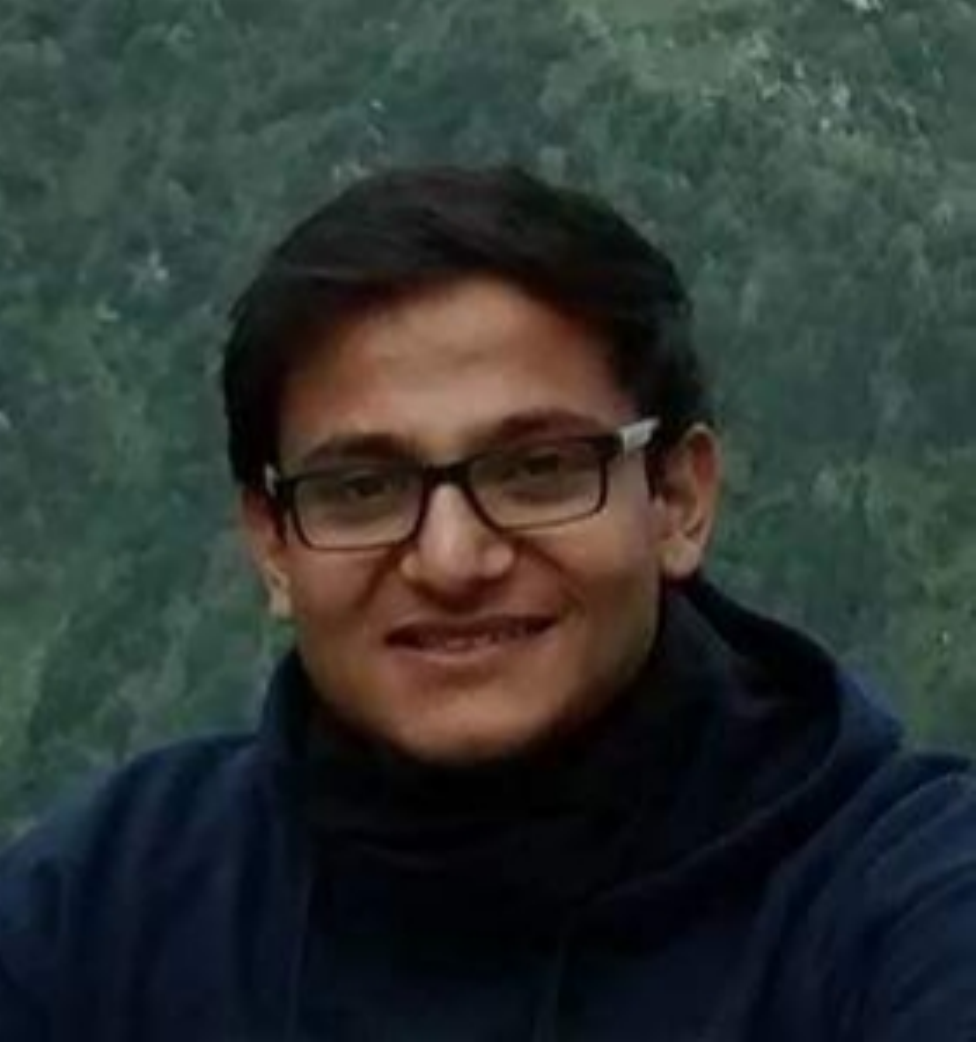}}]{Amogh Agrawal}
received the B.Tech degree in electrical engineering from the Indian Institute of Technology, Ropar, India in 2016. He was a research intern at University of Ulm, Germany in 2015, under the DAAD (German Academic Exchange Service) Fellowship. He joined the Nanoelectronics research lab in 2016, and is currently pursuing his PhD degree at Purdue University under the guidance of Prof. Kaushik Roy. His primary research interests include modeling and simulation of spin devices for application in logic, memories and neuromorphic computing. He is also looking at digital and analog circuits for in-memory computing techniques using CMOS and beyond-CMOS memories. He was awarded the Directors Gold Medal for his all-round performance, and University Silver Medal for his academic achievements at IIT Ropar. He is a recipient of the Andrews Fellowship from Purdue University.
\end{IEEEbiography}

\begin{IEEEbiography}[{\includegraphics[width=1in,height=1.25in,clip,keepaspectratio]{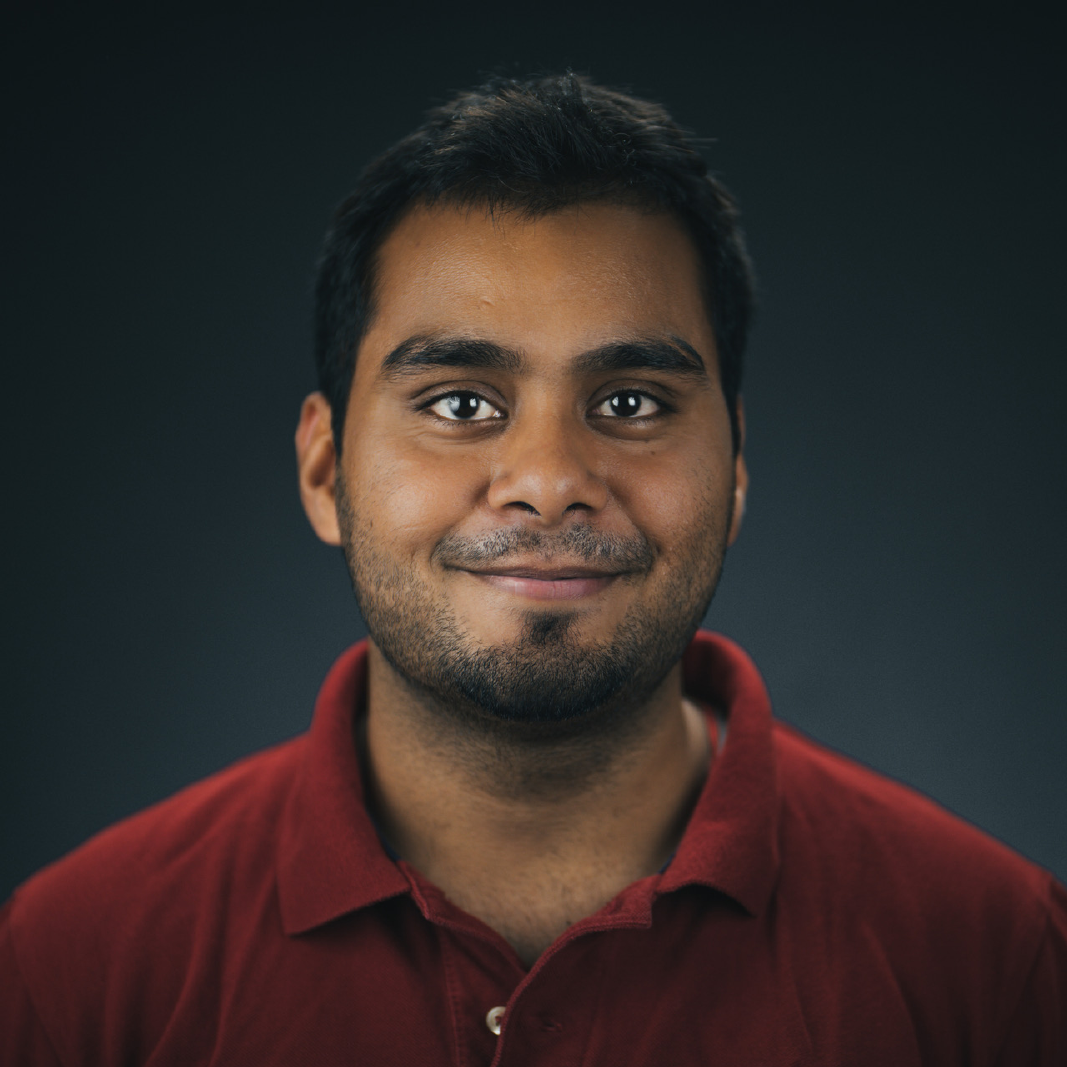}}]{Aayush Ankit} received the B.Tech. degree from Indian Institute of Technology (BHU), Varanasi in 2015. He was a summer intern and Mitacs Globalink Fellow at the University of Alberta, Canada in 2014. He has also worked as intern at HPE Labs, Palo Alto, CA and Intel Corporation, Hillsboro, OR in 2017. Currently, he is pursuing PhD degree in Electrical and Computer Engineering at Purdue University and is a Research Assistant to Prof. Kaushik Roy since Fall 2015. His primary research interests lie in architecture and algorithm design for Neuromorphic Computing.
\end{IEEEbiography}

\begin{IEEEbiography}[{\includegraphics[width=1in,height=1.25in,clip,keepaspectratio]{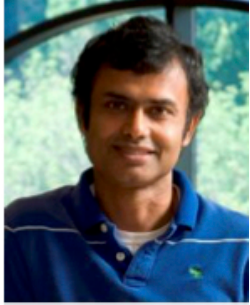}}]{Kaushik Roy}
received the BTech degree in electronics and electrical communications engineering from the Indian Institute of Technology, Kharagpur, India, and the PhD degree from the Department of Electrical and Computer Engineering, University of Illinois at Urbana-Champaign in 1990. He was with the Semiconductor Process and Design Center of Texas Instruments, Dallas, where he worked on FPGA architecture development and low-power circuit design. He joined the electrical and computer engineering faculty at Purdue University, West Lafayette, IN, in 1993, where he is currently Edward G. Tiedemann Jr. Distinguished Professor. His research interests include neuromorphic and cognitive computing, spintronics, device-circuit co-design for nano-scale Silicon and non-Silicon technologies, low-power electronics for portable computing and wireless communications, and new computing models enabled by emerging technologies. He has published more than 600 papers in refereed journals and conferences, holds 15 patents, graduated 70+ PhD students, and is coauthor of two books on Low Power CMOS VLSI Design (Wiley \& McGraw Hill). He received the US National Science Foundation Career Development Award in 1995, IBM faculty partnership award, ATT/Lucent Foundation award, 2005 SRC Technical Excellence Award, SRC Inventors Award, Purdue College of Engineering Research Excellence Award, Humboldt Research Award in 2010, 2010 IEEE Circuits and Systems Society Technical Achievement Award, Distinguished Alumnus Award from Indian Institute of Technology, Kharagpur, Fulbright Nehru Distinguished Chair, and Best Paper Awards at 1997 International Test Conference, IEEE 2000 International Symposium on Quality of IC Design, 2003 IEEE Latin American Test Workshop, 2003 IEEE Nano, 2004 IEEE International Conference on Computer Design, 2006 IEEE/ACM International Symposium on Low Power Electronics \& Design, and 2005 IEEE Circuits and System Society Outstanding Young Author Award (Chris Kim), 2006 IEEE Transactions on VLSI Systems Best Paper Award, 2012 ACM/IEEE International Symposium on Low Power Electronics and Design Best Paper Award, 2013 IEEE Transactions on VLSI Best Paper Award. He was a Purdue University Faculty scholar (1998-2003). He was a Research Visionary board member of Motorola Labs (2002) and held the M.K. Gandhi Distinguished Visiting faculty at Indian Institute of Technology (Bombay). He has been in the editorial board of IEEE Design and Test, IEEE Transactions on Circuits and Systems, IEEE Transactions on VLSI Systems, and IEEE Transactions on Electron Devices. He was the guest editor for Special Issue on Low-Power VLSI in the IEEE Design and Test (1994) and IEEE Transactions on VLSI Systems (June 2000), IEE Proceedings—Computers and Digital Techniques (July 2002), and IEEE Journal on Emerging and Selected Topics in Circuits and Systems (2011). He is a fellow of the IEEE.
\end{IEEEbiography}

\end{document}